\documentclass[9pt,twocolumn,twoside]{opticajnl}
\journal{opticajournal} % use for journal or Optica Open submissions

\setboolean{shortarticle}{True}
\usepackage{lineno}
\title{Pulse-Duration Scaling of Ultrafast Laser-Induced Damage Threshold in Hybrid Gratings}

\author[1,*]{Ziyao Su}
\author[1,2,3]{Enam A. Chowdhury}

\affil[1]{Department of Materials Science and Engineering, The Ohio State University, 140W. 19th Avenue, Columbus, OH 43210, USA}
\affil[2]{Department of Electrical and Computer Engineering, The Ohio State University, 2015 Neil Ave, Columbus, OH 43210, USA}
\affil[3]{Department of Physics, The Ohio State University, 191 W Woodruff Ave, Columbus, OH 43210, USA}

\affil[*]{su.775@osu.edu}

\begin{abstract}
High damage threshold gratings are in demand worldwide as critical components for next generation ultrahigh intensity lasers. Here we investigate the pulse-duration dependence of ultrafast laser-induced damage thresholds (LIDT) in hybrid multilayer dielectric gratings, touted to combine superior performance properties of both metallic and multilayer dielectric (MLD) gratings, using a dynamic finite-difference time-domain model incorporated with linear and non-linear absorption models. Simulations agree with reported experimental LIDT values for three representative designs and predict scaling exponents which vary with pulse durations ranging from 10 to 500~fs. The results reveal strong dependence on both material bandgap and grating field distribution, providing guidance for designing high LIDT gratings.
\end{abstract}

\setboolean{displaycopyright}{false} % Do not include copyright or licensing information in submission.

\begin{document}

\maketitle

\textbf{Introduction.}
The advancement of petawatt (PW) laser systems places increasingly urgent demands on pulse-compression gratings, with broad spectral bandwidth, high diffraction efficiency, and high laser-induced damage threshold (LIDT)~\cite{Wenfei2021}. Metallic gratings~\cite{Poole2013,Papadopoulos2016} offer intrinsically wide bandwidth but suffer from relatively high absorption, limiting their damage performance. Multilayer dielectric (MLD) gratings exhibit low absorption and high reflectivity~\cite{Jerald2004,Cotel2005,Guan2013,Simin2021,Su2025}; however, their bandwidth is constrained by interference design and typically requires more layers, which increases fabrication complexity and dispersion~\cite{Steinmeyer2006}, to extend.
Hybrid grating architectures~\cite{Neauport2010_1,Guan2014,Xu2019}, combining metallic and dielectric features, provide a promising approach to balance broadband operation with low loss and reduced dispersion. Such designs are particularly relevant for femtosecond PW systems, where both spectral bandwidth and peak-field resilience are critical~\cite{Wenfei2021}.

Establishing a reliable LIDT scaling is important for grating optimization. For a few picoseconds and longer pulses, damage thresholds generally follow the square-root scaling with pulse duration, consistent with thermally dominated mechanisms~\cite{Stuart1995,Stuart1996}. In contrast, the ultrafast regime is governed by electronic excitation and nonlinear ionization processes, and remains less systematically investigated, particularly in modeling, due to the difficulty of simulating picosecond-scale dynamics when complex ionization and collision processes are included.

Previously, we developed a dynamic model that couples a two-dimensional finite-difference time-domain (FDTD) field solver with multiple physical processes, including photoionization, impact ionization, electron collisions, electron heating, and refractive index modification for dielectric–laser interactions~\cite{Simin2022}, mainly under sub-100~fs irradiation, and recently extended it to metal–laser interactions with picosecond temporal capability~\cite{Su2026}. Here, we employ this framework to investigate the pulse-duration dependence of LIDT in hybrid gratings in the ultrafast regime. The model is benchmarked against reported experimental thresholds at representative pulse durations and subsequently extended to the 10--1000~fs range. The predicted scaling behavior shows good agreement with available measurements, providing insight into ultrashort-pulse damage trends relevant for next-generation PW laser systems.

\begin{figure}[ht!]
\centering\includegraphics[width=\linewidth]{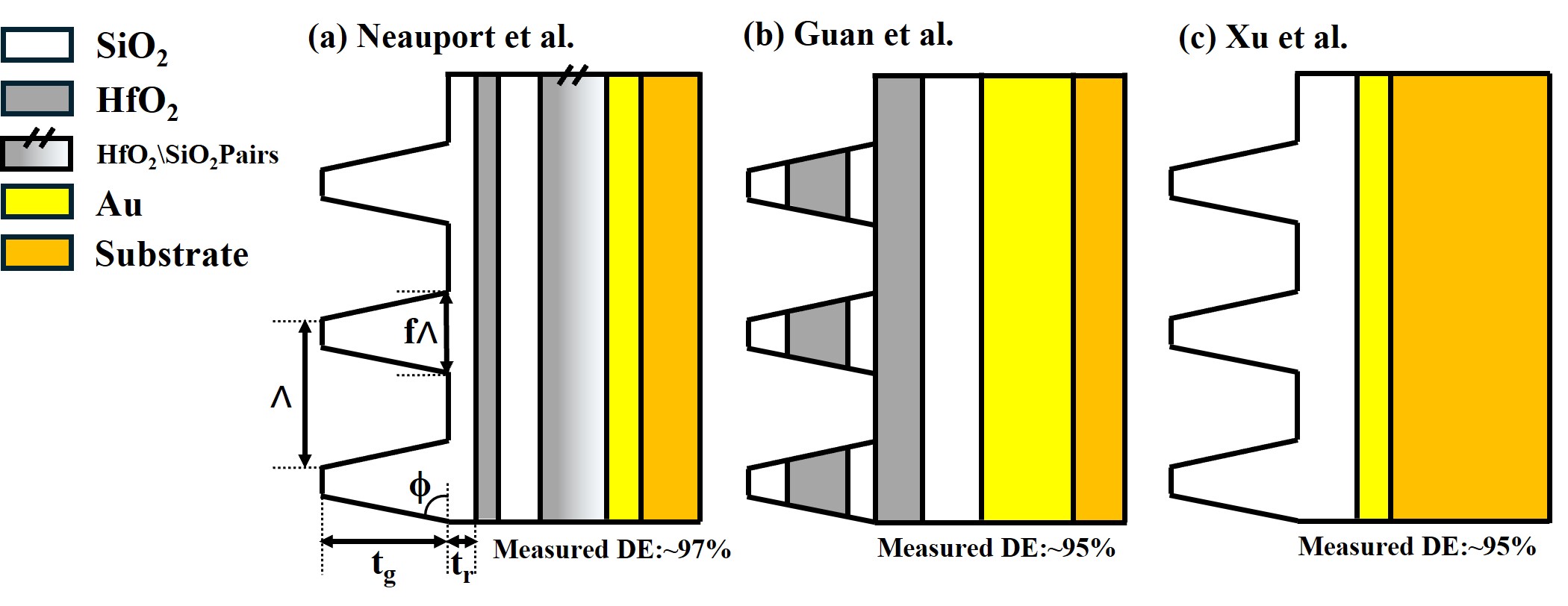}
\caption{Schematic illustrations of hybrid gratings: (a) Néauport \textit{et al.} (DE $\sim$ 97\%)~\cite{Neauport2010_1,Neauport2010_2}, (b) Guan \textit{et al.} (DE $\sim$ 95\%)~\cite{Guan2014}, and (c) Xu \textit{et al.} (DE $\sim$ 95\%)~\cite{Xu2019}. White indicates SiO$_2$, gray HfO$_2$, yellow Au, and orange the substrate. The three HfO$_2$/SiO$_2$ layer pairs are omitted in (a) for clarity. $\Lambda$ denotes the grating period, $f$ the duty cycle at the pillar bottom, $t_g$ the pillar depth, $t_r$ the residual thickness, and $\phi$ the sidewall angle.}
\label{fig:setup}
\end{figure}

\textbf{Simulation Set-up.}
Three representative dielectric-gold hybrid grating designs, selected based on fabrication complexity, are modeled over pulse durations ranging from 10--500~fs, with total simulation times extending to the ps scale, as shown in Fig.~\ref{fig:setup} with their reported measured peak diffraction efficiency (DE). Néauport \textit{et al.}~\cite{Neauport2010_1,Neauport2010_2} modified a SiO$_2$/HfO$_2$ MLD grating by replacing several bottom layers with gold, while retaining a relatively complex multilayer structure. Guan \textit{et al.}~\cite{Guan2014} reduced the layer count to six, with an HfO$_2$ layer embedded within SiO$_2$ pillars. Xu \textit{et al.} design~\cite{Xu2019} consisted of only two layers, SiO$_2$ and gold, representing the simplest architecture among the three.
The layer thicknesses and detailed grating parameters used in simulations follow the original designs and are summarized in Table S1 of Supplement 1.

The Néauport \textit{et al.} grating is modeled at a wavelength of 1057~nm with an angle of incidence (AOI) of 77.2$^\circ$. The Guan \textit{et al.} and Xu \textit{et al.} designs are simulated at 800~nm with AOIs of 53$^\circ$ and 65$^\circ$, respectively. All simulations assume \textit{s}-polarized incidence.
The material parameters used in the simulations are summarized in Tables S2--S4 of Supplement 1. For the Xu \textit{et al.} design, we adopt the material parameters explicitly reported in Ref.~\cite{Xu2019}. For the other two designs, the parameters primarily follow those established in our previous works~\cite{Simin2022,Su2026}.

\begin{figure*}[ht]
\centering
\includegraphics[width=\linewidth]{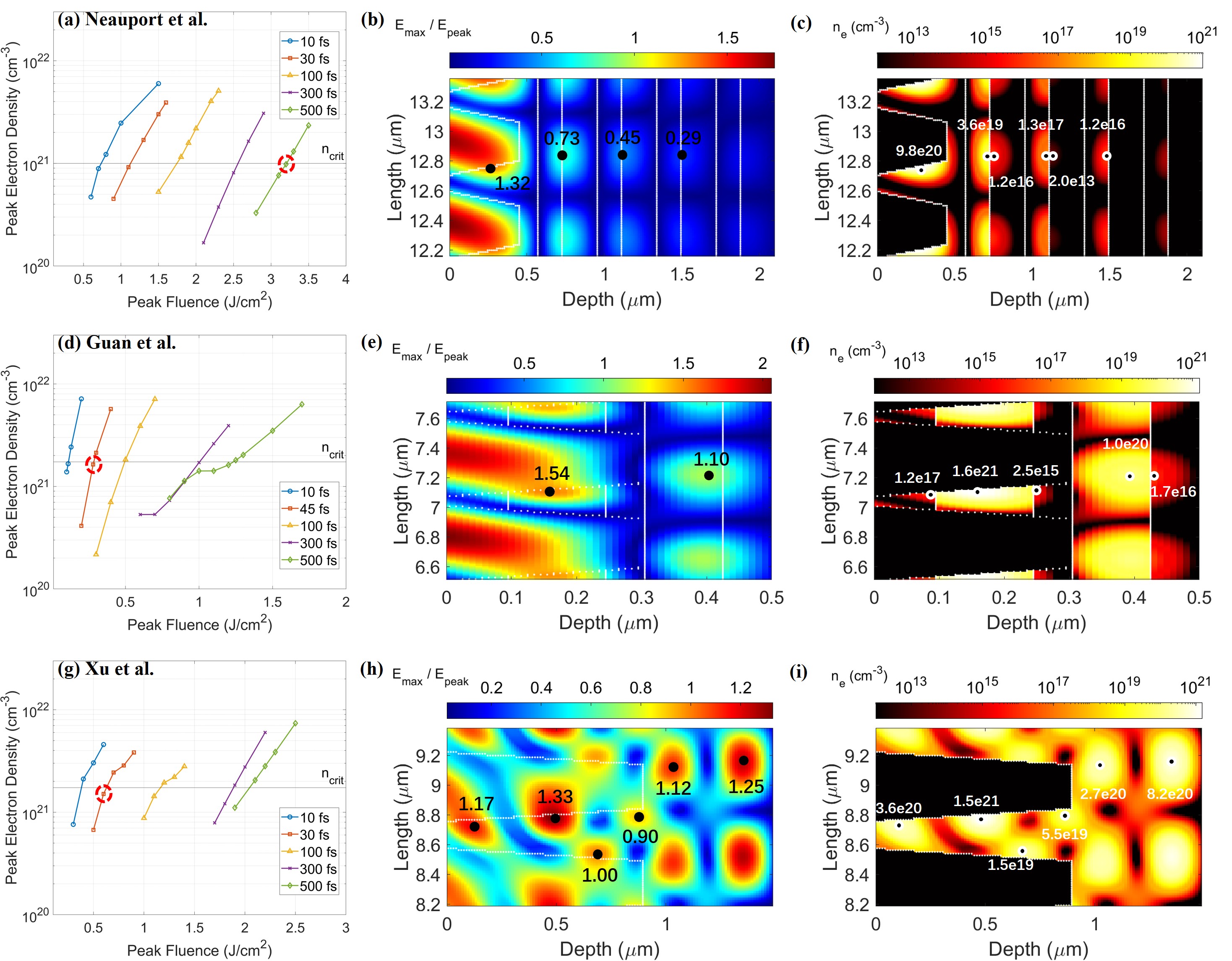}
\caption{Peak electron density as a function of irradiation fluence for pulse durations from 10--500~fs for the designs of (a) Néauport \textit{et al.}, (d) Guan \textit{et al.}, and (g) Xu \textit{et al.}. Panels (b,e,h) show the electric-field distributions and (c,f,i) the corresponding electron density distributions in the dielectric region at the fluence and pulse duration indicated by the red dashed circles in (a,d,g), corresponding to the experimental test durations reported in the original works: 500~fs for Néauport \textit{et al.}, 45~fs for Guan \textit{et al.}, and 30~fs for Xu \textit{et al.}.}
\label{fig:LIDT}
\end{figure*}

\textbf{Results.}
The peak electron densities at different fluences and pulse durations for the Néauport \textit{et al.} design are shown in Fig.~\ref{fig:LIDT}(a), indicating an increase in LIDT with pulse duration. 
At 3.2~J/cm$^{2}$ and 500~fs, the field enhancement distribution in Fig.~\ref{fig:LIDT}(b) shows the strongest localization at the sidewalls of the SiO$_2$ pillars, with a maximum enhancement of approximately 1.32, while the first HfO$_2$ layer exhibits a weaker enhancement of about 0.73. The corresponding electron density distribution in Fig.~\ref{fig:LIDT}(c) reveals a hotspot of $9.8\times10^{20}$~cm$^{-3}$ at the SiO$_2$ sidewalls, approaching the critical density ($1.0\times10^{21}$~cm$^{-3}$), indicating damage initiation at the pillars. The first HfO$_2$ layer reaches $3.6\times10^{19}$~cm$^{-3}$, and deeper dielectric layers exhibit electron densities more than three orders of magnitude lower. These results are consistent with experimentally reported damage initiation sites on the surface~\cite{Neauport2010_1,Neauport2010_2}.

For the Guan \textit{et al.} design~\cite{Guan2014}, Fig.~\ref{fig:LIDT}(d) similarly shows an increase in LIDT with pulse duration, although at fluences below 1~J/cm$^{2}$ the peak electron densities for the 300~fs and 500~fs cases reach comparable levels. At 0.28~J/cm$^{2}$ and 45~fs, the field enhancement distribution in Fig.~\ref{fig:LIDT}(e) indicates the strongest localization in the middle HfO$_2$ layer within the pillar, with a maximum enhancement of approximately 1.54, and a secondary hotspot in the second HfO$_2$ layer with an enhancement of about 1.10. The corresponding electron density distribution in Fig.~\ref{fig:LIDT}(f) shows peak electron densities of $1.6\times10^{21}$~cm$^{-3}$ and $1.0\times10^{20}$~cm$^{-3}$ in these regions, respectively, with the critical density being $1.7\times10^{21}$~cm$^{-3}$ at 800~nm. In contrast, the SiO$_2$ layers exhibit electron densities more than four orders of magnitude lower. These results indicate that damage initiation occurs in the middle HfO$_2$ layer within the pillar, followed by potential damage in the adjacent HfO$_2$ layer.

Similarly, for the Xu \textit{et al.} design~\cite{Xu2019}, Fig.~\ref{fig:LIDT}(g) also shows an increase in LIDT with pulse duration. Since this structure does not contain different dielectric layers, the interference pattern is less pronounced; however, multiple field hotspots appear along the pillar sidewalls and in the residual region. At 0.6~J/cm$^{2}$ and 30~fs, as shown in Fig.~\ref{fig:LIDT}(h), the strongest enhancement occurs near the middle of the pillar with a maximum value of approximately 1.33, while the second strongest hotspot appears near the bottom of the SiO$_2$ layer with an enhancement of about 1.25. The corresponding electron density distribution in Fig.~\ref{fig:LIDT}(i) shows peak electron densities of $1.5\times10^{21}$~cm$^{-3}$ and $8.2\times10^{20}$~cm$^{-3}$ at these locations, respectively, both approaching the critical density. These results indicate that damage initiation may occur at multiple locations along the top pillar sidewalls and near the base of the SiO$_2$ layer, which generally agree with the predictions by Xu \textit{et al.}~\cite{Xu2019}.

Additional electric field and electron density distributions for the three designs at different pulse durations and fluences are presented in Fig.~S1--S6 of Supplement 1, providing further comparison of the interaction dynamics.

\begin{figure}[ht!]
\centering\includegraphics[width=\linewidth]{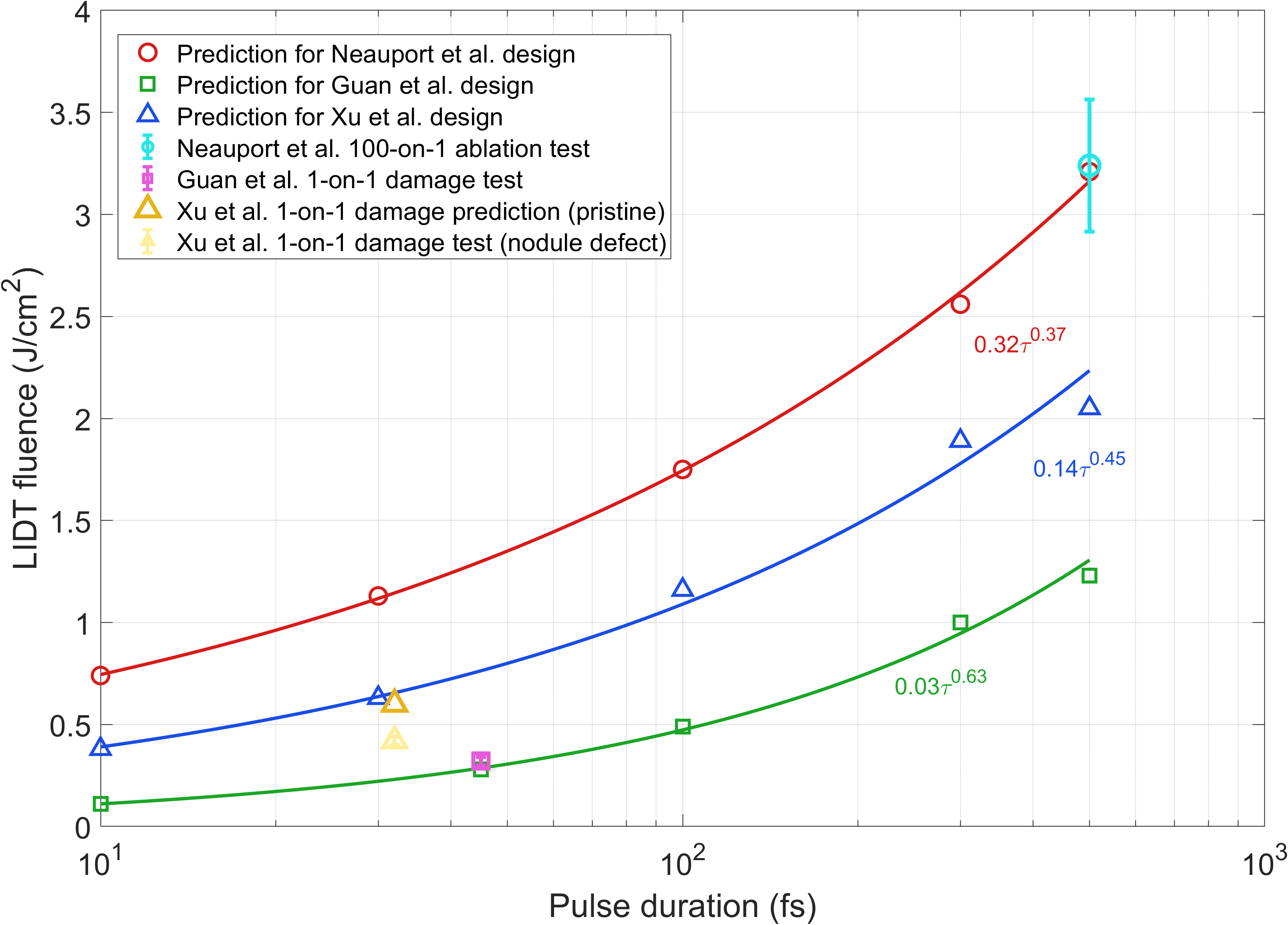}
\caption{Predicted relationship between LIDT and pulse duration below 500~fs for the grating designs of Néauport \textit{et al.} (red), Guan \textit{et al.} (green), and Xu \textit{et al.} (blue). The fitted curves are shown together with the originally reported LIDT values, marked using the same symbols.}
\label{fig:Tau_LIDT}
\end{figure}

The LIDT versus pulse duration for the three designs is shown in Fig.~\ref{fig:Tau_LIDT}, with the experimentally tested values indicated by markers reported in the original works. Uncertainty values are not explicitly reported in the work of Néauport \textit{et al.}. However, based on the measurement procedure described in their earlier study~\cite{Neauport2007}, the uncertainty is approximately $\sim2\%$ for a single site, with LIDT variations up to $\sim10\%$ across different testing locations. For both Guan \textit{et al.} and Xu \textit{et al.}, the measurement uncertainty is reported to be 0.02~J/cm$^2$ in the earlier work~\cite{Kong2017}. Our simulated results show good agreement with the reported LIDT values for all three designs at their tested pulse durations.

\textbf{Discussion.}
In Fig.~\ref{fig:LIDT}(d), an overlap of the peak electron density at 300~fs and 500~fs in the HfO$_2$ layer is observed. 
This behavior arises because the photoionization rate decreases slightly at the fluence corresponding to a peak electric field amplitude of about $6\times10^{9}$~V/m, as shown in Fig.~S7(a) in Supplementary 1. 
A comparison with the impact-ionization-off case for the Guan design is also presented in Fig.~S7(b) to isolate the contribution of photoionization. 
The results show that, at the same fluence, the 300~fs pulse generates slightly more electrons than the 500~fs pulse.

Additional features of the electron-density distribution can be observed at shorter durations. For the design of Néauport \textit{et al.}, the first HfO$_2$ layer can accumulate a relatively high electron density even when the hotspot in the pillar or surface layer reaches the critical density, as shown in Fig.~S4(a-2), Fig.~S5(a-2), and Fig.~\ref{fig:LIDT}(c). In these regions the electron density ranges from about half of the critical density to roughly two orders of magnitude lower, suggesting that these layers may also become susceptible to subsequent damage once breakdown is initiated at the primary hotspot. 
A similar behavior is observed for the design of Xu \textit{et al.}, where the hotspot near the Au layer reaches the critical density simultaneously with the sidewall hotspot for the 10~fs pulse, but decreases to about half of the critical density for longer pulse durations. 
These observations suggest that additional damage sites may become more likely in these locations for shorter pulses in grating structures.

Furthermore, the spatial distribution of the electron density appears more diffuse at shorter pulse durations, whereas clearer and more localized hotspots are observed for longer pulses. This behavior may be related to the broader spectral bandwidth of shorter pulses, which can lead to a less localized field distribution. 
Since mechanical effects are not included in the present model, the predicted LIDT for the 10~fs case may be somewhat overestimated, as the higher transient pressures generated in multiple regions could exceed the tensile strength of the material.

The relationship between LIDT of bulk dielectrics and pulse duration follows an empirical power law $\tau^{\alpha}$, in which heat diffusion dominates and yields $\alpha$ of about 0.5 for pulses longer than 20~ps~\cite{Stuart1995,Stuart1996}. 
Other works suggest $\alpha$ for bulk and MLD mirrors ranges between 0.2 to 0.8 under ns pulses~\cite{Jack1990,Garnov1993}. 
Reports~\cite{Mero2005,Deziel2021} further showed that fitting the scaling law for pulse durations from 30~fs to 1000~fs for bulk dielectrics yields $\alpha$ around 0.2 to 0.4. 
Recent work tested MLD gratings from 1~ps to 30~ps and reported $\alpha$ of about 0.22~\cite{Alessi2015}, while dielectric mirrors under sub-100~fs irradiation showed $\alpha$ near 0.25~\cite{Kong2012,Kong2013}. 
Our work investigates the duration dependence of hybrid gratings over 10~fs to 500~fs and shows $\alpha$ to range from 0.37 to 0.63, deviating from the 0.5 scaling observed for longer pulses, as shown in Fig.~\ref{fig:Tau_LIDT}.

We observe a decreasing trend of LIDT as the pulse duration is reduced. 
For the designs of Néauport \textit{et al.} and Xu \textit{et al.}, where damage initiates in the SiO$_2$ layers, $\alpha$ remains below 0.5, with values between 0.37 and 0.45. 
In contrast, for the Guan \textit{et al.} design, where damage initiates in the HfO$_2$ layer, $\alpha \approx 0.63$. 
The smaller bandgap of HfO$_2$ facilitates photoionization, allowing the electron density to reach the critical level more rapidly. 
As a result, breakdown is primarily governed by photoionization seeding, and the contribution of impact ionization becomes less dominant in determining the threshold. This behavior is illustrated in Fig.~S7(b), where impact ionization increases the electron density by only about a factor of two for the Guan \textit{et al.} design at 300~fs and 500~fs at fluences that reach the LIDT. 
In contrast, our earlier work~\cite{Su2026} shows that for the Néauport \textit{et al.} design, impact ionization can increase the electron density by nearly two orders of magnitude at 500~fs, indicating the stronger influences of impact ionization in the SiO$_2$ layers than in the HfO$_2$ layers when damage initiates.
Therefore, for the Guan \textit{et al.} design, the damage process remains more dependent on the cumulative generation of carriers during the pulse, leading to a stronger pulse-duration dependence and a larger $\alpha$ of 0.63.

Overall, the hybrid grating structures studied here below 1~ps exhibit slightly larger $\alpha$ values compared with previously reported $\alpha$ ranging from 0.2 to 0.4~\cite{Mero2005}.
One possible reason is that the local field distribution within the grating pillars is inherently more complex than in bulk dielectrics or planar MLD coatings, where the field enhancement is mainly governed by a relatively stable one-dimensional interference pattern along the layer stack. 
In the grating geometry, however, the hotspot is determined by a two-dimensional field distribution within the pillar structure. 
As free carriers are generated during the pulse, the resulting refractive index change can further modify the local field enhancement and redistribute the hotspot within the pillar region. 
Consequently, the electric field driving the ionization process evolves through a coupled interaction between plasma generation and field redistribution. 
This dynamic feedback alters the carrier-generation dynamics and can lead to a stronger dependence of the damage threshold on pulse duration, resulting in possible larger $\alpha$ in the hybrid grating structures.

Therefore, the scaling exponent $\alpha$ is determined by both the material properties, such as bandgap, and the structural field distribution within the grating, while the Au layer mainly influences the absolute threshold level through Ohmic absorption, which increases local energy deposition within the structure.

\textbf{Conclusion.}
The simulations reveal that electron density in dielectric layers beneath the surface layer or grating pillars can become significantly enhanced at shorter pulse durations, suggesting potential secondary damage locations within the multilayer structure. 
In addition, the spatial electron-density distribution becomes more diffuse for very short pulses, indicating that broadband ultrafast excitation may redistribute the field and carrier generation over a wider region.

Overall, our modeling shows good agreement with the experimentally reported LIDT values for three representative hybrid grating designs and predicts the pulse-duration dependence of the damage threshold for sub-picosecond irradiation. 
The fitted scaling exponent ranges from 0.37 to 0.63 for pulse durations between 10~fs and 500~fs, which is slightly higher than that typically reported for bulk dielectrics and planar MLD coatings in this regime, corresponding to a decreasing LIDT of hybrid gratings as the pulse duration shortens. 
The results further indicate that the scaling behavior depends on both material properties, such as bandgap, and the structural field distribution within the grating pillars, suggesting that establishing a universal scaling law for the ultrafast regime across different materials and structures remains challenging.

These findings provide useful guidance for predicting and optimizing the LIDT of hybrid grating optics used in PW-class laser systems for applications such as laser-driven secondary sources and particle accelerators. 
Future work will incorporate electron–lattice coupling through an extended two-temperature model, enabling simulations in the picosecond regime and allowing a more complete evaluation of the pulse-duration scaling of LIDT.

\begin{backmatter}
\bmsection{Funding} This work was partially supported by the DOE ARDAP award\# DE-SC0025577 and AFOSR award \# FA9550-25-1-0297.

\bmsection{Acknowledgment} This work used resources from the Ohio Supercomputer Center.

\bmsection{Disclosures} The authors declare no conflicts of interest.

\bmsection{Data availability} Data underlying the results presented in this paper are not publicly available at this time but may be obtained from the authors upon reasonable request.

\bmsection{Supplemental document}
See Supplement 1 for supporting content.
\end{backmatter}

% Bibliography
\bibliography{references}
% Full bibliography added automatically for Optics Letters submissions; the following line will simply be ignored if submitting to other journals.
% Note that this extra page will not count against page length
%\bibliographyfullrefs{references}

%Manual citation list
%\begin{thebibliography}{1}
%\bibitem{Zhang:14}
%Y.~Zhang, S.~Qiao, L.~Sun, Q.~W. Shi, W.~Huang, %L.~Li, and Z.~Yang,
 % \enquote{Photoinduced active terahertz metamaterials with nanostructured
  %vanadium dioxide film deposited by sol-gel method,} Opt. Express \textbf{22},
  %11070--11078 (2014).
%\end{thebibliography}

\end{document}